\documentclass[12pt]{article}
 \usepackage[cp1251]{inputenc} %- для WiN
 \usepackage[english, russian]{babel} %- для рус-англ. переноса
 \usepackage{amsmath, latexsym, amssymb, bm, array, graphics, eucal,
 amsfonts,}%mathscr
 \makeatletter
 \renewcommand{\@biblabel}[1]{#1.\hfill}
 \newcommand{\diag}{\rm \diag\, }

 \renewcommand{\Re}{\mathop{\rm Re}}
 \renewcommand{\Im}{\mathop{\rm Im\,}}

\renewcommand{\div}{\mathop{\rm div}}
\newcommand{\mc}[1]{\mathcal{#1}}
 \makeatother
 
\usepackage[dvips]{graphicx}
\usepackage[flushleft,small]{caption2} % чтобы в подписях к рисункам

\begin{document}
 \thispagestyle{empty}
 \large
 \renewcommand{\abstractname}{\,}
 \renewcommand{\refname}{\begin{center}REFERENCES\end{center}}
\newcommand{\E}{\mc{E}}
 \makeatother

\begin{center}
\bf Magnetic Susceptibility and Landau Diamagnetism of Quantum
Collisional Maxwellian Plasmas
\end{center}
\begin{center}
  \bf A. V. Latyshev\footnote{$avlatyshev@mail.ru$} and
  A. A. Yushkanov\footnote{$yushkanov@inbox.ru$}
\end{center}\medskip

\begin{center}
{\it Faculty of Physics and Mathematics,\\ Moscow State Regional
University, 105005,\\ Moscow, Radio str., 10--A}
\end{center}\medskip

\begin{abstract}
With the use of correct expression of the electric
conductivity of quantum collisional plasmas
( A. V. Latyshev and  A. A. Yushkanov,
{\it Transverse electrical conductivity of a quantum collisional
plasma in the Mermin approach}. - Theor. and Math. Phys., {\bf
175}:(1), 559--569 (2013)) the kinetic description of a magnetic
susceptibility is obtained and
the formula for calculation of Landau diamagnetism is deduced.

\medskip

electron gas
\end{abstract}

\begin{center}
\bf  Introduction
\end{center}

Magnetisation of electron gas in a weak magnetic fields
compounds of two independent parts (see, for example, \cite {Landau5}):
from the paramagnetic magne\-ti\-sa\-tion connected
with own (spin) magnetic
momentum of elect\-rons ({\it Pauli's para\-mag\-netism}, W. Pauli, 1927)
and from the diamagnetic mag\-ne\-ti\-sation connected with
quantization of
orbital movement of elect\-rons in a magnetic field ({\it
Landau diamagnetism}, L. D. Landau, 1930).

Landau diamagnetism  was considered till now for a gas of the free
elect\-rons. It has been thus shown, that together with original
approach develo\-ped by Landau, expression for diamagnetism of electron
gas can be obtained on the basis of the kinetic approach
\cite {Silin}.

The kinetic method gives opportunity to calculate the trans\-verse
die\-lect\-ric permeability.  On the basis of this quantity its possible
to obtain
the value of the diamagnetic response.

However such calculations till now
were carried out only for collisional\-less case. The matter is that
correct expression for the transverse dielectric
permeability of quantum plasma existed till
 now only for gas of the free
electrons. Expression known till now for the transverse dielectric
perme\-abi\-lity in  a collisional case gave incorrect transition to
the classical case \cite {Kliewer}. So this expression  were accordingly
incorrect.

Central result from \cite{Datta} connects the mean orbital
magnetic moment, a thermodynamic property, with the electrical
resistivity, which characterizes transport properties of
material. In this work
was discussed the important problem of  dissipation (collisions)
influence on  Landau diamagnetism. The analysis of this problem
is given with
use of exact expression of transverse conductivity of quantum plasma.

In work \cite{Kumar} is shown that a classical system of charged
particles moving on a finite but unbounded surface (of a sphere)
has a nonzero orbital diamagnetic moment which can be large.
Here is considered a non-degenerate system with the degeneracy
temperarure much smaller than the room temperature, as in the
case of a doped high-mobility semiconductor.

In work \cite{LY}  for the first time the expression   for
the quantum transverse dielectric
permeability of collisional plasma has been derived. The
obtained in \cite{LY} expression for
trans\-ver\-se dielectric permeability satisfies
to the necessary requirements of com\-pa\-ti\-bility.

In the present work for the first time  with use of correct
expression for the transverse conductivity \cite{LY} the
kinetic description of a magnetic susceptibility of quantum
collisional Maxwellian plasmas is given. The formula for
calculation of Landau
diamagnetism for Maxwellian collisional plasmas is deduced.

The graphic analysis of a magnetic susceptibility  and
comparison of a magnetic susceptibility Maxwellian and
degenerate plasmas is made.

In work \cite{7} the kinetic description has been considered
magnetic susceptibility and Landau diamagnetism of the quantum
collisional Maxwellian plasmas.

\begin{center}
  \bf 2. Magnetic susceptibility of quantum Maxwellian plasmas
\end{center}

Magnetization vector $\mathbf{M}$ of electron plasma
is connected with current density $\mathbf{j}$ by the following
expression \cite{Landau8}
$$
{\bf j}=c\, {\rm rot}\,\mathbf{M},
$$
where $c$ is the light velocity.

Magnetization vector $\mathbf{M}$ and a magnetic field
strength
$\mathbf{H}=\rm rot \mathbf{A}$ are connected by the expression
$$
{\bf M}=\chi\,{\bf H}=\chi\,{\rm rot}\,{\bf A},
$$
where $\chi$ is the magnetic susceptibility, $\mathbf{A}$ is the
vector potential.

From these two equalities for current density we have
$$
\mathbf{j}={c}\, {\rm rot}\,\mathbf{M}=
c\,\chi {\rm rot}\, \big({\rm rot}\,\mathbf{A}\big)=
c\,\chi\, \big[{\bf \nabla}({\bf \nabla}\cdot{\bf A})-
{\mathbf{\triangle}}\mathbf{A}\big].
$$

Here $\Delta$ is the Laplace operator.

Let the scalar potential is equal to zero.
Vector potential we take ortho\-gonal
to the direction of a wave vector $\mathbf{q}$
($\mathbf{q}\mathbf{A}=0$) in the form of a  harmonic wave
$$
\mathbf{A}(\mathbf{r},t)=\mathbf{A}_0
e^{i(\mathbf{q} \mathbf{r}-\omega t)}.
$$

Such vector field is solenoidal
$$
\div \mathbf{A}=\nabla\mathbf{A}=0.
$$

Hence, for current density we receive equality
$$
{\bf j}=-c\,\chi \Delta \mathbf{A}=c\,\chi\,q^2 \mathbf{A}.
$$

On the other hand, the connection of electric field  $\mathbf{E}$ and vector
potential $\mathbf{A}$ is as follows
$$
\mathbf{E}=-\dfrac{1}{c}\dfrac{\partial \mathbf{A}}{\partial t}=
\dfrac{i\omega}{c}\mathbf{A}.
$$
It is leads to the relation
$$
\mathbf{j}=\sigma_{tr}\mathbf{E}=\sigma_{tr}\dfrac{i\omega}{c}
\mathbf{A},
$$
where $\sigma_{tr}$ is the transverse electric conductivity.

For our case from (1.1) and (1.2) we obtain
following expression for the magnetic susceptibility
$$
\chi=\dfrac{i\omega}{c^2 q^2}\sigma_{tr}.
\eqno{(1.1)}
$$

Expression of transversal conductivity of Maxwellian collisional
plasmas it is defined by the general formula \cite{LY}:
$$
\sigma_{tr}({\bf q},\omega,\nu)=\sigma_0\dfrac{i \nu}{\omega}\Big(1+
\dfrac{\omega B({\bf q},\omega+i\nu)+i \nu B({\bf q},0)}
{\omega+i \nu}\Big),
\eqno{(1.2)}
$$
where $\sigma_0$ is the static conductivity,
$\sigma_0={e^2N}/{m\nu}$, $N$ is the concentration (number density)
of plasmas particles, $e$ and $m$ is the electron charge and mass,
$\nu$ is the effective collisional frequency of plasmas particles,
$$
B({\bf q},0)=\dfrac{\hbar^2}{8\pi^3mN}\int
\dfrac{f_{\bf k}-f_{\bf k-q}}
{\E_{\bf k}-\E_{\bf k-q}}{\bf k}_\perp^2d^3k,
$$

$$
B({\bf q},\omega+i\nu)=\dfrac{\hbar^2}{8\pi^3mN}
\int \dfrac{f_{\bf k}-f_{\bf k-q}}
{\E_{\bf k}-\E_{\bf k-q}-\hbar(\omega+i \nu)}{\bf k}_\perp^2d^3k,
$$
$$
f_{\bf k}=\dfrac{4\pi^{3/2}\hbar^3}{m^3v_T^3}
\exp\Big(-\dfrac{\E_{{\bf k}}}{k_BT}\Big),\qquad
\E_{\bf k}=\dfrac{\hbar^2{\bf k}^2}{2m},
$$
$$
\E_T=\dfrac{mv_T^2}{2}=\dfrac{\hbar^2k_T^2}{2m}=k_BT,
$$
$\E_{\bf k}$ is the electrons energy,
$\E_T$ is the heat electrons energy, $k_B$ is the Boltzmann's
constant,
$v_T=1/\sqrt{\beta}$ is the heat electrons velocity,
$\beta=m/2k_BT$,
$\hbar$ is the Planck's constant,

$$
{\bf k}_\perp^2={\bf k}^2-\Big(\dfrac{{\bf kq}}{q}\Big)^2.
$$

According to (1.1) and (1.2) magnetic susceptibility of the quantum
collisional Maxwellian plasmas it is equal
$$
\chi({\bf q},\omega,\nu)=-\dfrac{e^2N}{mc^2q^2}\Big(1+
\dfrac{\omega B({\bf q},\omega+i\nu)+i \nu B({\bf q},0)}
{\omega+i \nu}\Big).
\eqno{(1.3)}
$$

From the formula (1.3) it is visible, that at $\omega=0$ frequency
of collisions plasma particles $ \nu $ drops out of the formula (1.3).
Hence, the magnetic susceptibility in a static limit does not depend from
frequencies of collisions of plasma and the following form also has:

$$
\chi({\bf q},0,\nu)=-\dfrac{e^2N}{mc^2q^2}\Big[1+
\dfrac{\hbar^2}{8\pi^3mN}\int \dfrac{f_{\bf k}-f_{\bf k-q}}
{\E_{\bf k}-\E_{\bf k-q}}{\bf k}_\perp^2d^3k\Big].
\eqno{(1.4)}
$$

From expression (1.3) it is visible, that a magnetic susceptibility in
collisionless quantum Maxwellian plasma is equal:
$$
\chi({\bf q},\omega,0)=-\dfrac{e^2N}{mc^2q^2}\Big[1+
\dfrac{\hbar^2}{8\pi^3mN}\int \dfrac{f_{\bf k}-f_{\bf k-q}}
{\E_{\bf k}-\E_{\bf k-q}-\hbar\omega}{\bf k}_\perp^2d^3k\Big].
\eqno{(1.5)}
$$
At $ \omega\to  0$ the formula (1.5) passes in the formula (1.4).

Let's deduce the formula for calculation of a magnetic susceptibility
of quantum collisional Maxwellian plasmas.

After obvious linear replacement of variables the formula for integral
$B({\bf q}, \omega+i \nu)$ will be transformed to the form
$$
B({\bf q}, \omega+i \nu)=\dfrac{\hbar^2}{8\pi^3mN}\times $$$$ \times
\int \dfrac{(\E_{\bf k+q}+\E_{\bf k-q}-2\E_{\bf k})
f_{\bf k}{\bf k}_\perp^2d^3k}
{[\E_{\bf k}-\E_{\bf k-q}-\hbar(\omega+i \nu)]
[\E_{\bf k+q}-\E_{\bf k}-\hbar(\omega+i \nu)]}.
\eqno{(1.6)}
$$

Let's enter dimensionless variables
$$
z=\dfrac{\omega+i \nu}{k_Tv_T}=x+iy, \quad
x=\dfrac{\omega}{k_Tv_T}, \quad y=\dfrac{\nu}{k_Tv_T}, \quad
Q=\dfrac{q}{k_T}.
$$
Let's pass to integration on the vector ${\bf K} = {\bf k}/k_T$,
where $k_T=p_T/\hbar=mv_T/\hbar $ is the thermal wave number.
Vectors $ {\bf K, k} $ we will direct along an axis $x $, believing $ {\bf
K} =K_x (1,0,0) $ and $ {\bf k} =k (1,0,0) $.

Then
$$
\E_{{\bf k}}=\dfrac{\hbar^2k_T^2}{2m}K^2=\E_TK^2,
$$
$$
\E_{\bf k}-\E_{\bf k-q}-\hbar(\omega+i \nu)=2\E_TQ\Big(K_x-\dfrac{z}{Q}-
\dfrac{Q}{2}\Big),
$$
$$
\E_{\bf k+q}-\E_{\bf k}-\hbar(\omega+i \nu)=2\E_TQ\Big(K_x-\dfrac{z}{Q}+
\dfrac{Q}{2}\Big),
$$
$$
\E_{\bf k+q}+\E_{\bf k-q}-2\E_{\bf k}=2\E_TQ^2,
$$
$$
f_{{\bf k}}=\dfrac{4\pi^{3/2}}{k_T^3}
\exp\Big(-\dfrac{\E_{{\bf k}}}{\E_T}\Big)=
\dfrac{4\pi^{3/2}}{k_T^3}\exp(-K^2).
$$

The magnetic susceptibility  in dimensionless variables is equal
$$
\chi(Q,x,y)=-\dfrac{e^2N}{mc^2k_T^2Q^2}
\Big(1+\dfrac{x}{z}B(Q,z)+\dfrac{iy}{z}B(Q,0)\Big).
\eqno{(1.7)}
$$
Here according to (1.6)
$$
B(Q,z)=\dfrac{1}{2\pi^{3/2}}\int \dfrac{e^{-K^2}K_\perp^2 d^k}
{(K_x-z/Q)^2-(Q/2)^2},
$$
$$
B(Q,0)=\dfrac{1}{2\pi^{3/2}}\int \dfrac{e^{-K^2}K_\perp^2 d^k}
{K_x^2-(Q/2)^2}.
$$

The internal double integral is equal
$$
\int\limits_{-\infty}^{\infty}\int\limits_{-\infty}^{\infty}
e^{-K_y^2-K_z^2}(K_y^2+K_z^2)dK_ydK_z=\pi.
$$

Hence, last two integrals are reduced to the one-dimensional
integrals
$$
B(Q,z)=\dfrac{1}{2\sqrt{\pi}}\int\limits_{-\infty}^{\infty}
\dfrac{e^{-\tau^2}d\tau}{(\tau-z/Q)^2-(Q/2)^2},
$$
$$
B(Q,0)=\dfrac{1}{2\sqrt{\pi}}\int\limits_{-\infty}^{\infty}
\dfrac{e^{-\tau^2}d\tau}{\tau^2-(Q/2)^2}.
$$

\begin{center}
\bf 2. Landau diamagnetism of quantum Maxwellian
collisionless plasmas
\end{center}

Landau diamagnetism  in collisionless  plasma is usually
 defined as a magnetic susceptibility in a static limit
for a homogeneous external magnetic field.
Thus the diamagnetism value can be found by means of (1.1)
through two  non-commutative limits
$$
\chi_L=\lim\limits_{q\to 0}\Big[\lim\limits_{\omega\to 0}^{}
\chi(q,\omega,\nu=0)\Big].
\eqno{(2.1)}
$$

Into collisionless  plasma
this expression (2.1) should lead to the known formula of
Landau's diamagnetism of quantum Maxwellian plasmas \cite{Anselm}
$$
\chi_L=-\dfrac{e^2 N}{6m c^2k_T^2}=-\dfrac{e^2Nk_BT\hbar^2}{3m^4c^2}.
\eqno{(2.2)}
$$

So the relative magnetic susceptibility by means of (2.2)
is equal:
$$
\dfrac{\chi(Q,x,y)}{\chi_L}=\dfrac{6}{Q^2}
\Big(1+\dfrac{x}{z}B(Q,z)+\dfrac{iy}{z}B(Q,0)\Big).
\eqno{(2.3)}
$$

Let's deduce by means of expression (2.3) formula
for Landau dia\-mag\-netism. At $z=0$ from the formula (2.3)
for magnetic susceptibility the quantum
collisionless Maxwellian plasmas we obtain the following expression
$$
\dfrac{\chi(Q)}{\chi_L}=\dfrac{6}{Q^2}\Big[1+
\dfrac{1}{2\sqrt{\pi}Q}\int\limits_{-\infty}^{\infty}
\dfrac{e^{-\tau^2}-e^{-(\tau-Q)^2}}{\tau-Q/2}d\tau\Big],
\eqno{(2.4)}
$$
or
$$
\dfrac{\chi(Q)}{\chi_L}=\dfrac{6}{Q^2}\Big[1+
\dfrac{1}{2\sqrt{\pi}}\int\limits_{-\infty}^{\infty}
\dfrac{e^{-\tau^2}d\tau}{\tau^2-(Q/2)^2}\Big].
$$

Let's prove with use (2.4), that
$$
\dfrac{\chi(0)}{\chi_L}=\lim\limits_{Q\to 0}\dfrac{6}{Q^2}\Big[1-
\dfrac{1}{2\sqrt{\pi}Q}\int\limits_{-\infty}^{\infty}
\dfrac{e^{-(\tau-Q)^2-e^{-\tau^2}}}{\tau-Q/2}d\tau\Big]=1.
\eqno{(2.5)}
$$
This relation also will prove equality (2.2).
Really, we will spread out function
$$
\varphi(Q)=\dfrac{e^{-(\tau-Q)^2}-e^{-\tau^2}}{\tau-Q}
$$
in series on degrees $Q $ near to a point $Q=0$
$$
\varphi(Q)=2e^{-\tau^2}Q+2\tau e^{-\tau^2}Q^2+
\Big(\dfrac{4}{3}\tau^2-1\Big)Q^3+ \cdots.
$$

Now according to (2.5) we obtain
$$
\dfrac{\chi(0)}{\chi_L}=\lim\limits_{Q\to 0}\dfrac{6}{Q^2}\Bigg\{1-
\dfrac{1}{2\sqrt{\pi}}\int\limits_{-\infty}^{\infty}e^{-\tau^2}
\Big[2+2\tau e^{-\tau^2}Q+\Big(\dfrac{4}{3}\tau^2-1\Big)Q^2+
\cdots\Big]d\tau\Bigg\}=
$$
$$
=-\dfrac{3}{Q^2}\dfrac{1}{\sqrt{\pi}}\int\limits_{-\infty}^{\infty}e^{-\tau^2}
\Big(\dfrac{4}{3}\tau^2-1\Big)d\tau=1,
$$
as was to be shown.

\begin{center}\bf
  3. Аnalysis of results
\end{center}

For graphic research of the magnetic susceptibility we will be
to use the formula (2.4).

On fig. 1 comparison of a magnetic susceptibility
of Maxwellian and degenerate plasmas as functions of the dimensionless
wave number for a case of collisionless plasmas in
static limit is presented.

From fig. 1 it is clear, that there is a value
of dimensionless wave number $Q_*$, in which values
of magnetic susceptibility of degenerate and Maxwellian plasmas
are equal each other. So at 0$\leqslant Q <Q_* \; (Q> Q_*)$
values  of a magnetic susceptibility of degenerate plasma it is more
(less) than values of the magnetic susceptibility of
Maxwellian plasmas.

From fig. 1 it is obvious, that into quantum collisionless plasma
($ \nu=0$) in static limit ($ \omega=0$) the magnetic
susceptibility is function of wave number, monotonously decreasing to zero
both for degenerate, and for Maxwellan plasmas.

On fig. 2 and 3 dependence of magnetic susceptibility
of collisionless plasmas as from wave number (fig. 2), and
from dimensionless frequency of oscillations of an
electromagnetic field $x $ (fig. 3) is presented.

From fig. 2 it is obvious,
that the magnetic susceptibility is monotonously decreasing
function of wave number at all values of frequency of oscillations
of electromagnetic field. Thus for all $x <1$
($ \omega <\omega_p $) values of a magnetic susceptibility that
more than the values of frequency of oscilluations of the
electromagnetic fields is more.

\begin{figure}[h]
\begin{flushleft}
\includegraphics[width=15.0cm, height=9cm]{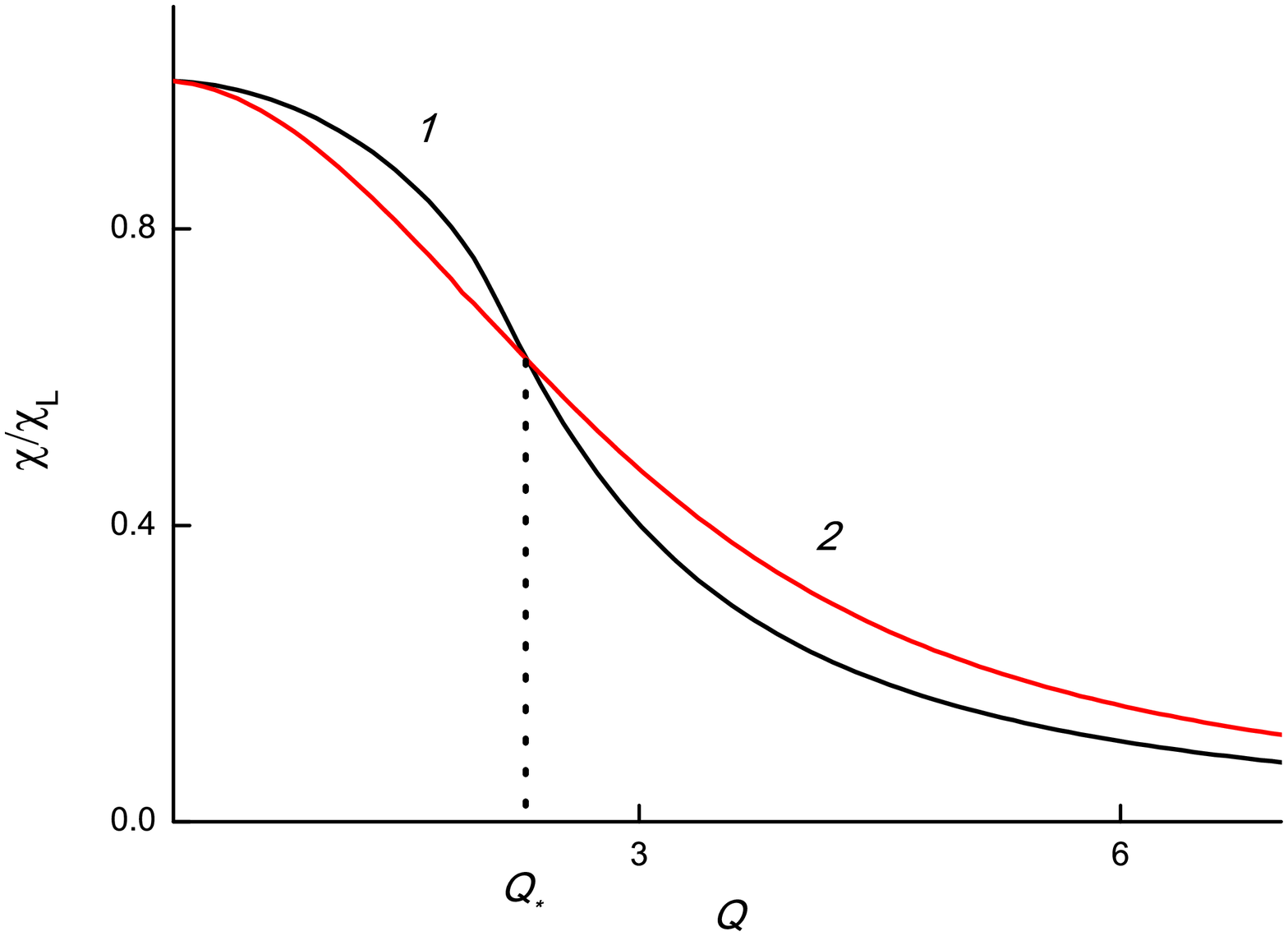}
{Fig 1. Magnetic susceptibility in static limit ($\omega=0$) of quantum
collisionless  ($\nu=0$). Curve 1 corresponds to degenerate
plasma, curve 2 corresponds to Maxwellian plasma.
}\label{rateIII}
\end{flushleft}
\end{figure}

From fig. 3 it is clear, that a magnetic susceptibility as function
of oscillation frequencies of electromagnetic field has
the maximum near to frequency $ \omega=Q\omega_p $
and with growth $Q $ moves to the right.

On fig. 4 and 5 dependence of real (fig. 4)
and imaginary (fig. 5) parts of magnetic susceptibility from quantity
of dimensionless frequency of oscillations of electromagnetic
field in the case $Q=0.5$ is presented.

From fig. 4 it is clear, that the real part has a maximum,
which is displaced to the right with growth of frequency
of collisions of plasma particles.
Irrespective of frequency of collisions
of particles of plasma with
growth of frequency of fluctuations of an electromagnetic field the
quanttity of the real part of a magnetic susceptibility leaves
from above on the asymptotics
$$
\lim\limits_{x\to 0}\Re\Big(\dfrac{\chi(Q, x, y)}{\chi_L}\Big) =
\dfrac{6}{Q^2}.
$$

Not resulting necessary graphics we will inform, that with
decrising values of wave number a maximum of a magnetic susceptibility
moves to the left and becomes sharp at small values of frequency
collisions of particles of plasma. With growth of frequency of collisions
the maximum starts to smooth out and vanishes.

From fig. 5 it is obvious, that an imaginary part of
a magnetic susceptibility as
function of dimensionless frequency of oscillations
of an electromagnetic field
has a minimum. This minimum moves to the left with  growth frequency
collisions of particles of plasma. With growth of the dimensionless
frequencies of oscillations of an electromagnetic
field an imaginary part of the magnetic
susceptibilities leaves from below on the asymptotics
$ \Im(\chi/\chi_L) =0$. We will notice, that a minimum
of an imaginary part not
vanishes with growth as frequencies of collisions of
particles of plasma, and
dimensionless wave number.

Let's notice, that the less frequency of collisions of
particles of plasma, the
it is more than value of the real and imaginary parts of the magnetic
susceptibilities.

\begin{center}
  \bf 4. Conclusion
\end{center}

In the present work the kinetic description of magnetic
susceptibilities of quantum collisional Maxwellian
plasmas with use before deduced correct formulas for
transversal electric conductivity of quantum plasma is given.

Influence of collisions of particles of plasma on the magnetic
suscep\-ti\-bi\-lity is found out.
Thereby the answer to a question on influence  dissipa\-ti\-ons on
Landau diamagnetism put in work \cite{Datta} is given.
For collisionless plasmas with the help
the kinetic approach the known formula of Landau dia\-mag\-ne\-tism
is deduced.

\begin{figure}[h]
\begin{flushleft}
\includegraphics[width=15.0cm, height=10cm]{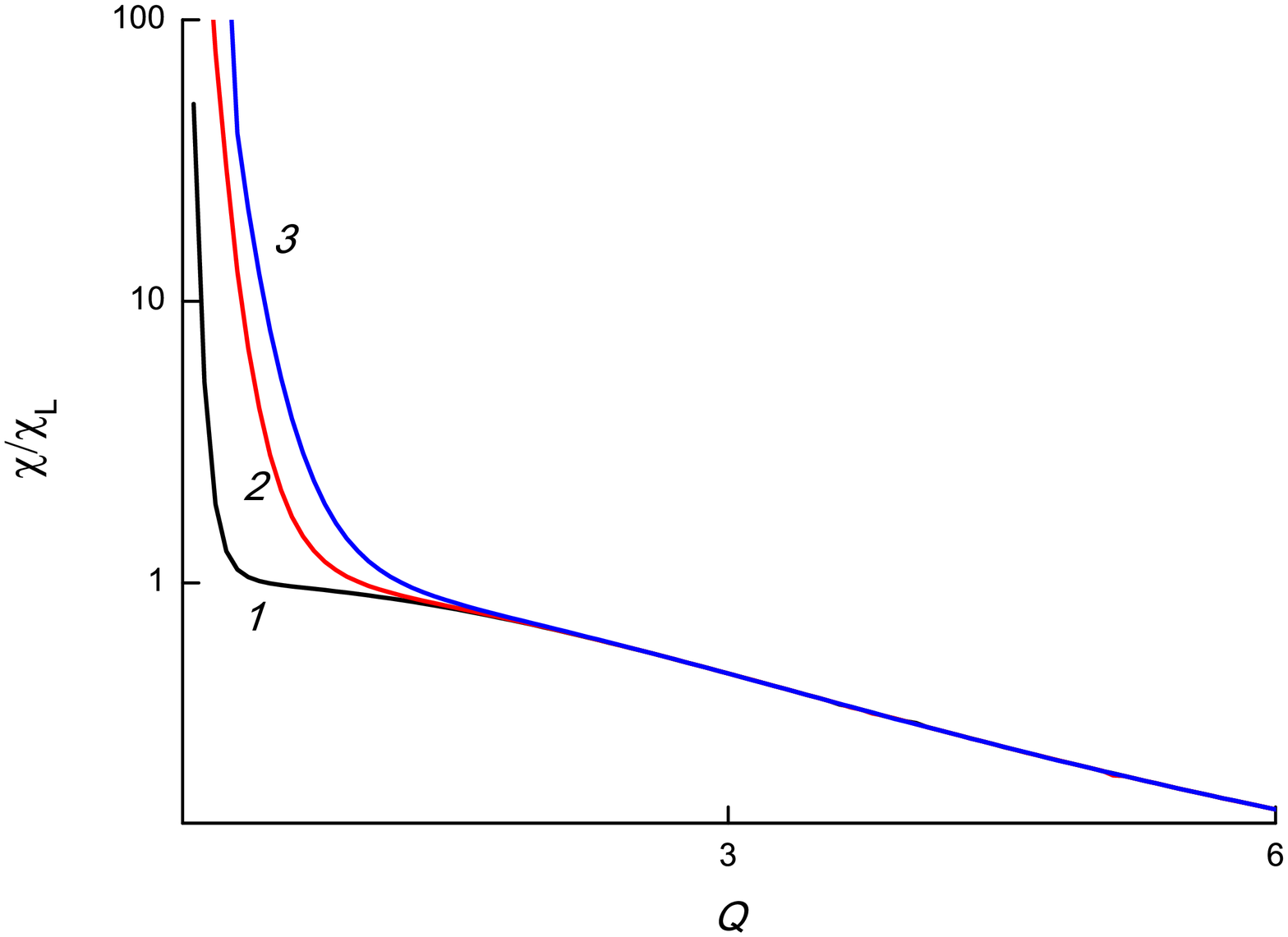}
{Fig 2. Magnetic susceptibility of collisionless plasma,
curves $1$,$2$ and $3$ correspond to parameter values
$x=0.01, 0.1$ and $x=0.2$.
}\label{rateIII}
\end{flushleft}
%\end{figure}
%\begin{figure}[h]
\begin{flushleft}
\includegraphics[width=15.0cm, height=10cm]{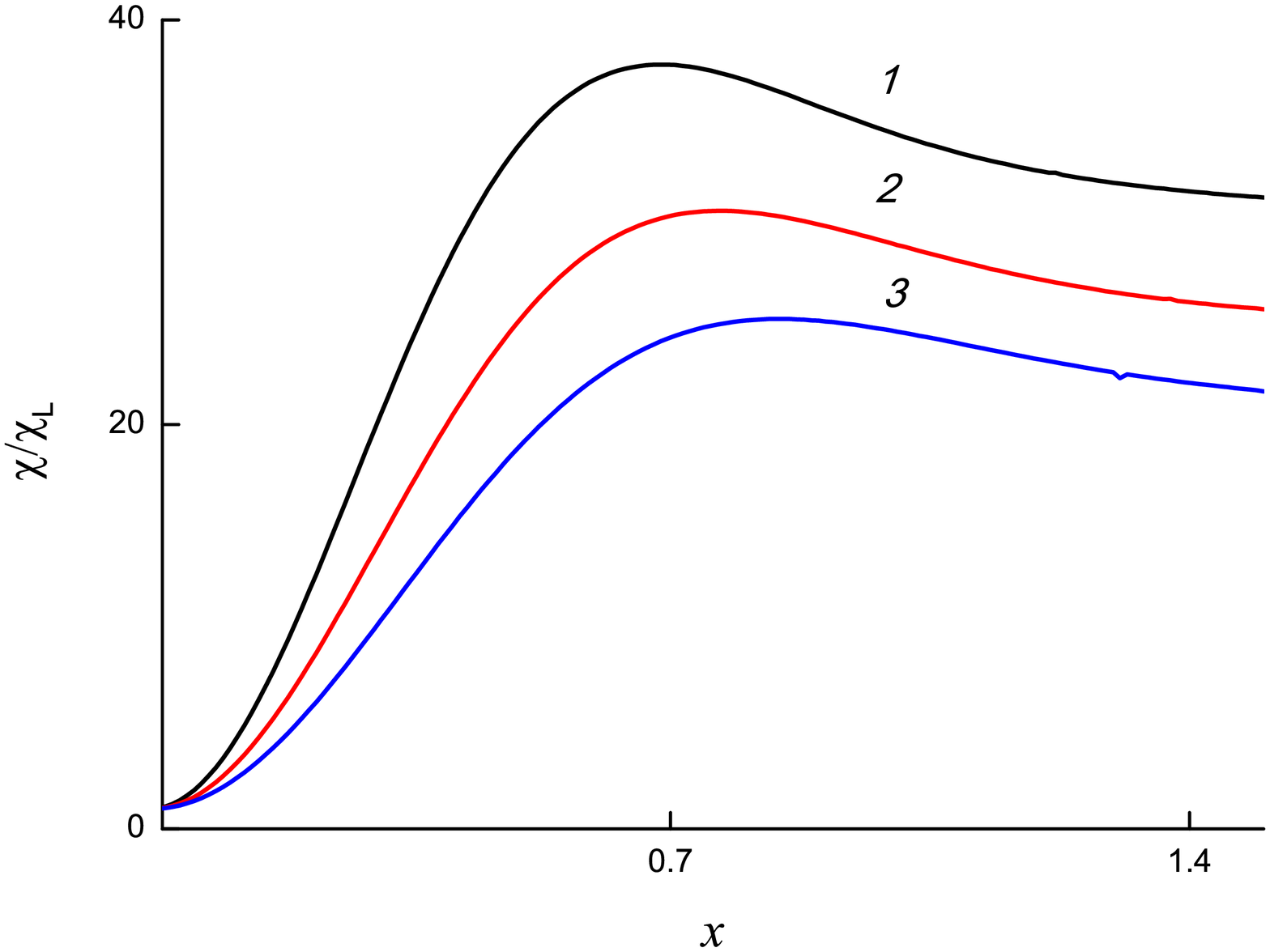}
{Fig 3. Magnetic susceptibility of collisionless plasma,
curves $1$, $2$, $3$ correspond to parameter values
$Q=0.45, 0.50, 0.55$.
}\label{rateIII}
\end{flushleft}
\end{figure}
\clearpage
\begin{figure}[h]
\begin{flushleft}
\includegraphics[width=15.0cm, height=10cm]{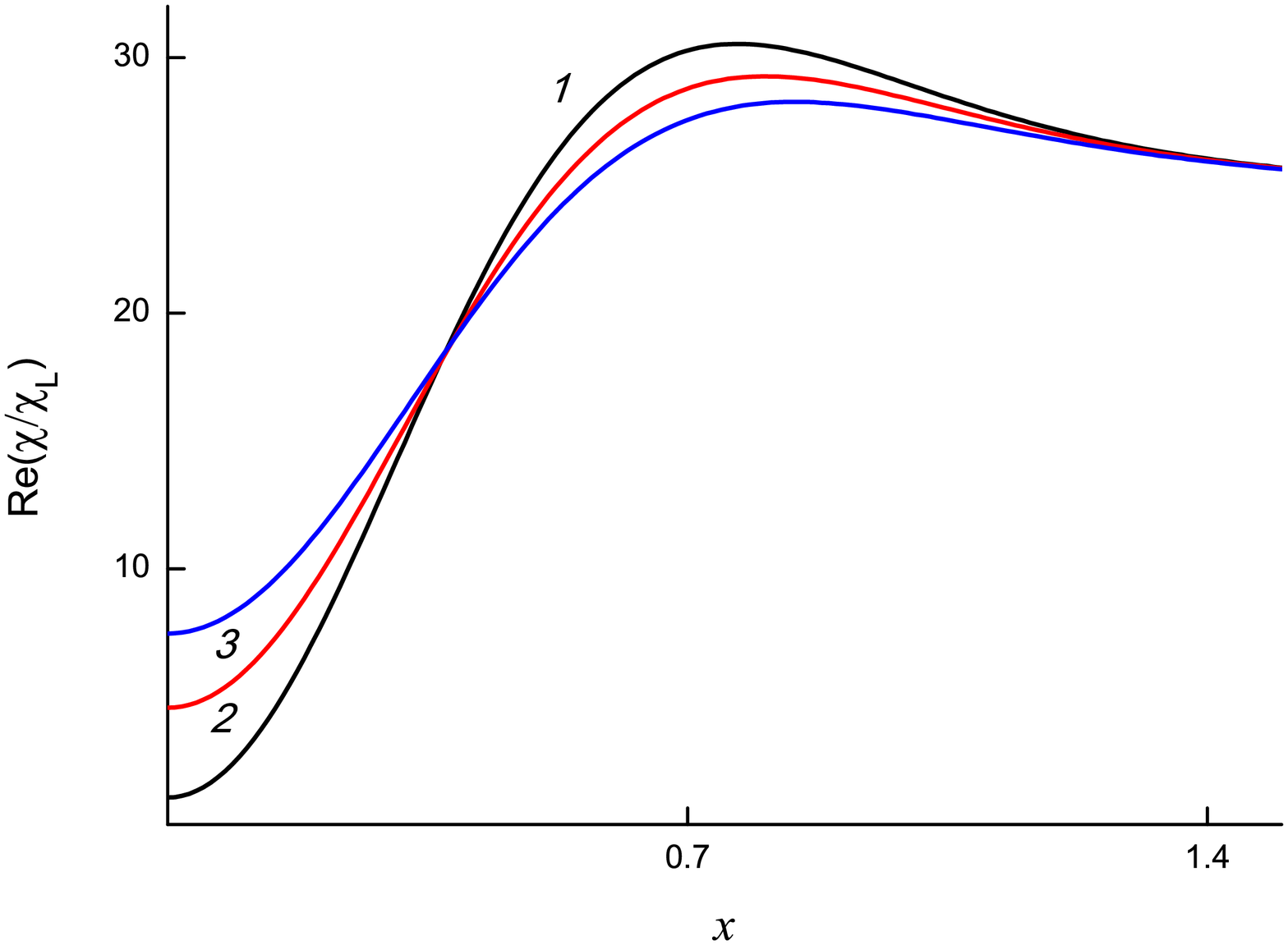}
{Fig 4. Real part of magnetic susceptibility; $Q=0.5$, curves
$1,2,3$ correspond to parameter values $y=0.001, 0.05, 0.1$.
}\label{rateIII}
\end{flushleft}
%\end{figure}
%\begin{figure}[h]
\begin{flushleft}
\includegraphics[width=15.0cm, height=10cm]{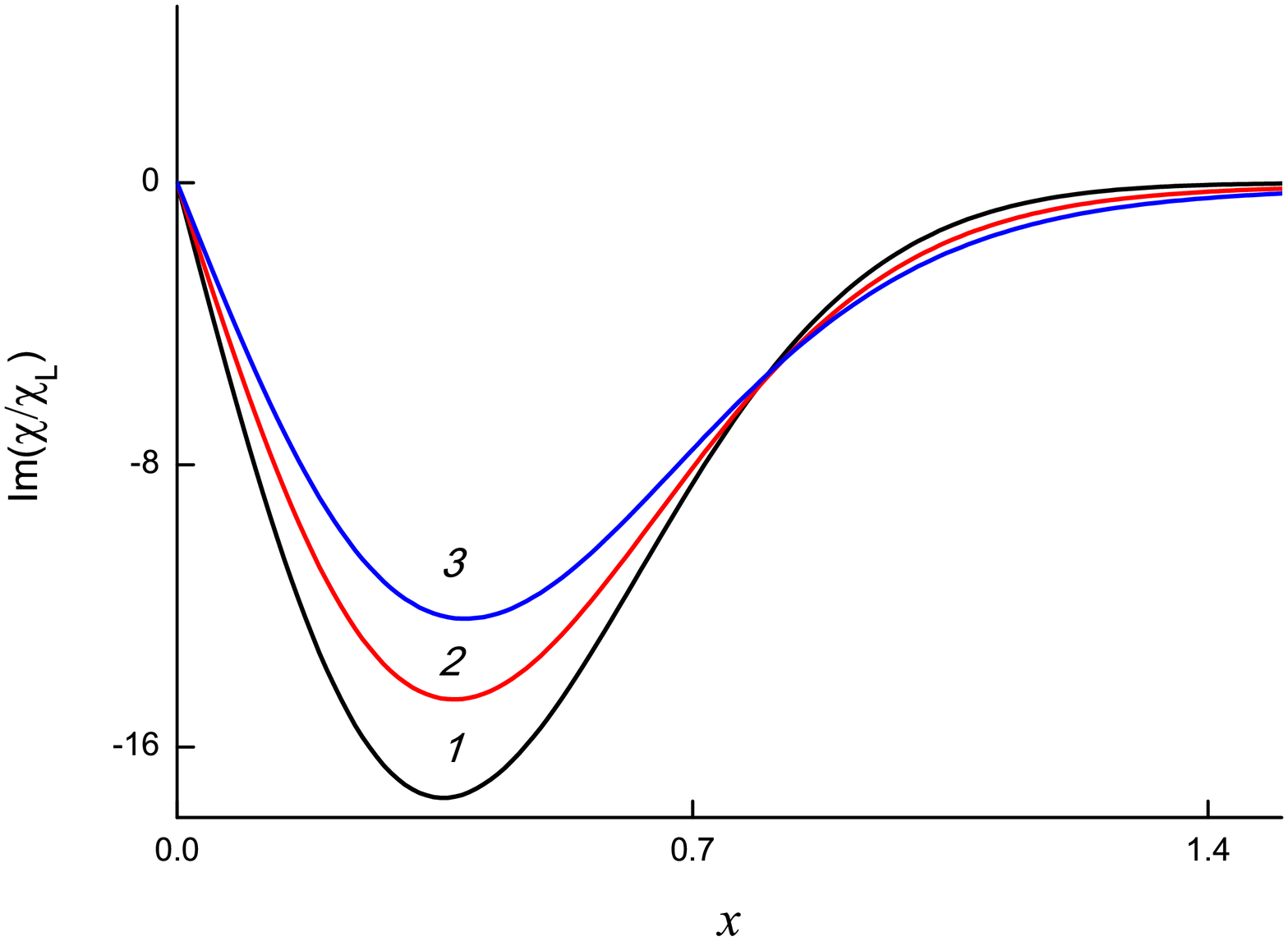}
{Fig 5. Imaginary part of magnetic susceptibility; $Q=0.5$, curves
$1,2,3$ correspond to parameter values $y=0.001, 0.05, 0.1$.
}\label{rateIII}
\end{flushleft}
\end{figure}
%\clearpage

\end{document}